\documentclass[preprint]{revtex4}
\usepackage{graphicx}
\usepackage{dcolumn}
\usepackage{bm}
\usepackage{amsmath}
\usepackage{amsfonts}
\usepackage{amssymb}
\begin{document}

\title{\textbf{ Frequency-dependent local interactions and low-energy effective
models from electronic structure calculations }}
\author{F. Aryasetiawan$^{a}$, M. Imada$^{b,c}$, A. Georges$^{d,e}$, G. Kotliar$^{f}$,
S. Biermann$^{d}$, and A. I. Lichtenstein$^{g}$}
\date{Jan 2004}
\begin{abstract}
We propose a systematic procedure for constructing
effective models of strongly correlated materials. The parameters, in
particular the on-site screened Coulomb interaction \emph{U}, are calculated
from first principles, using the GW approximation. We derive an expression for
the frequency-dependent $U(\omega)$ and show that its high frequency part has
significant influence on the spectral functions. We propose a scheme for
taking into account the energy dependence of $U(\omega)$, so that a model with
an energy- independent local interaction can still be used for low-energy properties.
\end{abstract}

\affiliation{$^a$Research Institute for Computational Sciences, AIST,
1-1-1 Umezono, Tsukuba Central 2, Ibaraki 305-8568, Japan}
%\author{M. Imada}
%
\affiliation
{$^b$Institute for Solid State Physics, University of Tokyo, Kashiwaoha,
Kashiwa, Chiba, 277-8581, Japan} \affiliation
{$^c$PRESTO, Japan Science and Technology Agency}
%\author{A. Georges}
%
\affiliation{$^d$Centre de Physique Theorique, Ecole Polytechnique,
91128 Palaiseau, France} \affiliation{$^e$LPT-ENS CNRS-UMR 8549, 24 Rue
Lhomond, 75231 Paris Cedex 05, France}
%\author{G. Kotliar}
%
\affiliation{$^f$Department of Physics and Astronomy,
Serin Physics Laboratory,
Rutgers University,
Piscataway
NJ 08854-8019
USA}
%\author{S. Biermann}
%\affiliation{Centre de Physique Theorique, Ecole Polytechnique,
%91128 Palaiseau, France}
%\author{A. I. Lichtenstein}
%
\affiliation{$^g$University of Nijmegen, NL-6525 ED, Nijmegen, The Netherlands}

\maketitle

\section{Introduction}

Lattice fermion models such as the Hubbard model or the Anderson impurity
model and their extensions have played a major role in studying electron
correlations in systems with strong onsite correlations. Despite the
widespread use of these models, little justification has been given in using
them. The models are postulated on the basis of physical intuition. In
particular, the models employ parameters, such as the famous Hubbard
interaction \emph{U,}\textbf{\ }which are normally adjusted to serve the given
problem. Without judicious choice of parameters, the model may yield
misleading results or in the worst case, the model itself is not sufficient to
describe the real system. One can define rigorously \ these concepts in the
path integral formulation of the many body problem by performing a partial
trace over the degrees of freedom that one wants to eliminate, and ignoring
the retardation in the interactions generated by this procedure. However, this
elimination of the degrees of freedom is very hard to perform for real
materials. It is therefore very desirable to figure out a systematic way of
constructing
%the Hubbard or Anderson impurity model 
%
low energy effective models with well defined parameters calculated from first
principles such that the model can quantitatively reproduce and predict
physical properties of interest of the corresponding real system, especially
when the correlation effects are crucial.

Another important issue that has not received sufficient attention is the role
of energy dependence of the screened local Coulomb interaction \emph{U.}
%Hubbard \emph{U.} 
%
Model studies investigating the importance of high-energy states in the
Hubbard model can be found in \cite{olle2,sun,florens}. A dynamic Hubbard
model has also been considered. \cite{hirsch}. In most cases however \emph{U}
is assumed to be static, but on the other hand we know that at high energy the
screening becomes weaker and eventually the interaction approaches the large
bare Coulomb value, which is an order of magnitude larger than the static
screened value. Of course the high-energy part of the Coulomb interaction has
in some way been down-folded into the Hubbard \emph{U} but it is not clear how
this downfolding is actually accomplished.

A number of authors have addressed the problem of determining the Hubbard
\emph{U} from first principles. One of the earliest works is the constrained
local density approximation (LDA) approach \cite{olle1,olle2} where the
Hubbard \emph{U} is calculated from the total energy variation with respect to
the occupation number of the localized orbitals. An approach based on the
random-phase approximation (RPA) was later introduced \cite{springer}, which
allows for the calculations of the matrix elements of the Hubbard \emph{U} and
its energy dependence. This was followed by a more refined approach for
calculating \emph{U}. \cite{kotani}. A yet different approach, computes the
matrix elements of the Coulomb interactions screened in real space
%, which then assume 
%
and assumes a Yukawa form to extract the Hubbard U and the other interactions
which determine the multiplet splittings. \cite{brooks}.

The purpose of the present work is to develop a precise formulation for a
systematic construction of effective models where the parameters are obtained
from realistic first-principles electronic structure calculations. In
particular, we concentrate on the calculation of the Hubbard \emph{U}%
\textbf{\ }and demonstrate the importance of its energy dependence.
%AG
%
%
We show that a static Hubbard hamiltonian, obtained from a naive construction
in which this energy dependence is simply neglected, fails even at low energy.
This static model can be appropriately modified however, by taking into
account the feedback of the high-energy part of \emph{U} into the one-particle
propagator.
%
%We also propose a scheme that takes into
%account the high-energy part of \emph{U}.
%
%
We illustrate our scheme in transition metals, concentrating on Ni as an
example, since it is a prototype system consisting of a narrow 3d band
embedded in a wide band. Furthermore, Ni is
%AG
%
%
one of the most problematic case from the viewpoint of the LDA.

\section{Theory}

Let us suppose that the bandstructure of a given solid can be separated into a
narrow band near the Fermi level and the rest, like, for example, in
transition metals or 4f metals. Our aim is to construct
%AG
%a lattice Hubbard model
%
%
an effective model which only includes the narrow 3d or 4f band. The effective
interaction between the 3d electrons in the Hubbard model can be formally
constructed as follows. We first divide the complete Hilbert space into the
Hubbard space $\{\psi_{d}\}$, consisting of the 3d states or the localized
states, and the rest.
%AG
%
%
The bare Green's function $G_{d}$, spanning the d-subspace is given by:
\begin{equation}
G_{d}(\mathbf{r,r}^{\prime};\omega)=\sum_{d}^{occ}\frac{\psi_{d}%
(\mathbf{r)}\psi_{d}^{\ast}(\mathbf{r}^{\prime})}{\omega-\varepsilon
_{d}-i0^{+}}+\sum_{d}^{unocc}\frac{\psi_{d}(\mathbf{r)}\psi_{d}^{\ast
}(\mathbf{r}^{\prime})}{\omega-\varepsilon_{d}+i0^{+}} \label{HubbardG}%
\end{equation}
%AG
% Maybe one should restore the sentence below ?
%Strictly speaking, $\{\psi_{d}\}$ should correspond to the Hartree
%wavefunctions.
%
%AG I rewrote a bit the paragraph on P, for clarity
%
%
%
Let $P$ be the total (bare) polarization, including the transitions between
all bands:
\begin{align}
P(\mathbf{r,r}^{\prime};\omega)  &  =\sum_{i}^{occ}\sum_{j}^{unocc}\psi
_{i}(\mathbf{r)}\psi_{i}^{\ast}(\mathbf{r}^{\prime})\psi_{j}^{\ast
}(\mathbf{r)}\psi_{j}(\mathbf{r}^{\prime})\nonumber\\
&  \times\left\{  \frac{1}{\omega-\varepsilon_{j}+\varepsilon_{i}+i0^{+}
}-\frac{1}{\omega+\varepsilon_{j}-\varepsilon_{i}-i0^{+}}\right\}  \label{P}%
\end{align}
$P$ can be divided into: $P=P_{d}+P_{r}$, in which $P_{d}$ includes only 3d to
3d transitions (i.e limiting the summations in (\ref{P}) to $i,j\in\{\psi
_{d}\}$), and $P_{r}$ be the rest of the polarization. The screened
interaction $W$ on the RPA level is given by
\begin{align}
W  &  =[1-vP]^{-1}v\nonumber\\
&  =[1-vP_{r}-vP_{d}]^{-1}v\nonumber\\
&  =[(1-vP_{r})\{1-(1-vP_{r})^{-1}vP_{d}\}]^{-1}v\nonumber\\
&  =\{1-(1-vP_{r})^{-1}vP_{d}\}^{-1}(1-vP_{r})^{-1}v\nonumber\\
&  =[1-W_{r}P_{d}]^{-1}W_{r} \label{W}%
\end{align}
where we have defined a screened interaction $W_{r}$ that does not include the
polarization from the 3d-3d transitions:
\begin{equation}
W_{r}(\omega)=[1-vP_{r}(\omega)]^{-1}v \label{Wr}%
\end{equation}
%AG
%
%
(we have not explicitly indicated spatial coordinates in this equation). The
\emph{identity} in (\ref{W}) explicitly shows that the interaction between the
3d electrons is given by a frequency-dependent interaction $W_{r}$. It fits
well with the usual physical argument that the remaining screening channels in
the Hubbard model associated with the 3d electrons, represented by the 3d-3d
polarization $P_{d},$ further screen $W_{r}$ to give the fully screened
interaction $W$.

%AG: I rewrote a bit the following paragraph, for clarity.
%
%
We now choose a basis of Wannier functions $\{\phi_{Rn}\}$, centered about
atomic positions R, corresponding to the 3d Bloch functions $\{\psi
_{\mathbf{k}n}\}$, and consider the matrix elements of the (partially
screened) frequency-dependent Coulomb interaction $W_{r}$:
\begin{equation}
U_{R_{1}nR_{2}n^{\prime},R_{3}mR_{4}m^{\prime}}(\tau-\tau^{\prime}%
)\doteqdot\int d^{3}rd^{3}r^{\prime}\phi_{R_{1}n}^{*}(\mathbf{r)}\phi
_{R_{2}n^{\prime}}(\mathbf{r})W_{r}(\mathbf{r,r}^{\prime};\tau-\tau^{\prime
})\phi_{R_{3}m}^{*}(\mathbf{r}^{\prime}\mathbf{)}\phi_{R_{4}m^{\prime}%
}(\mathbf{r}^{\prime}) \label{Uk}%
\end{equation}
We would like to obtain an effective model for the 3d degrees of freedom.
Because of the frequency dependence of the U's (corresponding to a retarded
interaction), this effective theory will not take a hamiltonian form. We can
however, write such a representation in the functional integral
formalism\cite{negele_orland} by considering the effective action for the 3d
degrees of freedom given by:
\begin{align}
S  &  =\int d\tau d\tau^{\prime}\left[  -\sum d_{Rn}^{\dagger}(\tau
){}_{Rn,R^{\prime}n^{\prime}}^{-1}(\tau-\tau^{\prime})d_{Rn^{\prime}}%
(\tau^{\prime})\right. \nonumber\\
&  \left.  +\frac{1}{2}\sum:d_{R_{1}n}^{\dagger}(\tau)d_{R_{2}n^{\prime}}%
(\tau):U_{R_{1}nR_{2}n^{\prime},R_{3}mR_{4}m^{\prime}}(\tau-\tau^{\prime
}):d_{R_{3}m}^{\dagger}(\tau^{\prime})d_{R_{4}m^{\prime}}(\tau^{\prime
}):\right]  \label{S}%
\end{align}
where $:d^{\dagger}d:$ denotes normal ordering, which accounts for the Hartree
term, and the summation is over repeated indices.
%AG
% I changed the paragraph below once more as compared to version 2.
%
%
When using a Wannier transformation which does not mix the d-subspace with
other bands, the Green's function ${}$ can be taken, to first approximation,
to be the bare Green's function $G^{0}_{dd}$ constructed from the Bloch
eigenvalues and eigenfunctions. If instead an LMTO formalism
\cite{andersen-lmto} is used, one should in principle obtain ${}$ from a
downfolding procedure onto the d-subspace, i.e perform a partial trace over
$s,p$ degrees of freedom (e.g to first order: ${}^{-1}=[G_{dd}]^{-1}%
-[G_{ds}]^{-1}G_{ss}[G_{sd}]^{-1}$).

In the following, we retain only the local components of the effective
interaction on the same atomic site. This is expected to be a reasonable
approximation because the 3d states are rather localized. The formalism may be
easily extended to include intersite Coulomb interactions if necessary. Hence,
we consider the frequency-dependent Hubbard interactions:
\begin{equation}
U_{nn^{\prime},mm^{\prime}}(\tau-\tau^{\prime})\doteqdot\int d^{3}%
rd^{3}r^{\prime}\phi_{n}^{\ast}(\mathbf{r})\phi_{n^{\prime}}(\mathbf{r}%
)W_{r}(\mathbf{r,r}^{\prime};\tau-\tau^{\prime})\phi_{m}^{\ast}(\mathbf{r}%
^{\prime})\phi_{m^{\prime}}(\mathbf{r}^{\prime}) \label{U}%
\end{equation}
with $\phi_{n}$ being the Wannier orbital for R=0. In order to illustrate the
procedure within the linear muffin-tin orbital (LMTO) basis set, we use
instead of the Wannier orbital the normalized function head of the LMTO
$\phi_{H}$ which is a solution to the radial Shr{\"o}dinger equation matching
to a Hankel function at zero energy at the atomic sphere boundary.

In this paper, we investigate the importance of the energy- dependence of
\emph{U}. Therefore, we shall compare the results obtained from (\ref{S}) with
those of a Hamiltonian approach in which one would construct a Hubbard model
with a \textit{static} interaction \emph{U}:
\begin{equation}
H=\sum_{Rn,R^{\prime}n^{\prime}} c^{\dagger}_{Rn} h_{Rn,R^{\prime}n^{\prime}}
c_{R^{\prime}n^{\prime}} +\frac{1}{2}\sum_{R,nn^{\prime},mm^{\prime}}
c^{\dagger}_{Rn}c_{Rn^{\prime}}U_{nn^{\prime},mm^{\prime}} c^{\dagger}_{Rm}
c_{Rm^{\prime}} \label{ham}%
\end{equation}
It seems natural to identify the static Hubbard $\emph{U}$ with the
(partially) screened local interaction in the low-frequency limit
$W_{r}(\omega=0)$.
%AG
% I explained the motivation above, so I removed this. In fact, we do not really
% discuss whether a different choice of the static U could be better, and this is not
% our main goal.
%
% but this
%assumption requires justification since we know that the true effective
%interaction should be energy dependent.
%
%
%
Note that the Hubbard model (\ref{ham}) has been constructed in the most naive
manner, by simply taking the quadratic part to be the d-block of the
non-interacting hamiltonian.

In order to compare the results obtained from the full dynamical U to those of
the static Hubbard model, we need to solve (\ref{S}) and (\ref{ham}) within
some consistent approximation scheme. In the following, we adopt a ``GW
universe'' where the exact self-energy for the solid is assumed to be given by
the GW approximation (GWA)~\cite{hedin,ferdi},
%AG
%
%
and where the GW approximation is also assumed to be a reliable tool in
solving the effective models (\ref{S}) and (\ref{ham}). This allows us to make
a proper comparison between the ``exact'' self-energy and the Hubbard model
self-energy. If the assumption of static \emph{U} is valid, the Hubbard
self-energy and the true self-energy (both within the GWA) should be close to
each other, at least for small energies. Or equivalently, the spectral
function for small energies should resemble that of the full one.%
%TCIMACRO{\FRAME{ftbpFU}{3.0753in}{2.2399in}{0pt}{\Qcb{respect to the
%\TEXTsymbol{\backslash}\$E\TEXTsymbol{\backslash}\_g\TEXTsymbol{\backslash}\$
%channel. Other channels give almost the same result.}}{\Qlb{ReWr}}%
%{rewr.eps}{\special{ language "Scientific Word";  type "GRAPHIC";
%width 3.0753in;  height 2.2399in;  depth 0pt;  original-width 9.436in;
%original-height 6.8571in;  cropleft "0";  croptop "1";  cropright "1";
%cropbottom "0";  filename 'ReWr.eps';file-properties "XNPEU";}}}%
%BeginExpansion
\begin{figure}
[ptb]
\begin{center}
\includegraphics[
height=2.2399in,
width=3.0753in
]%
{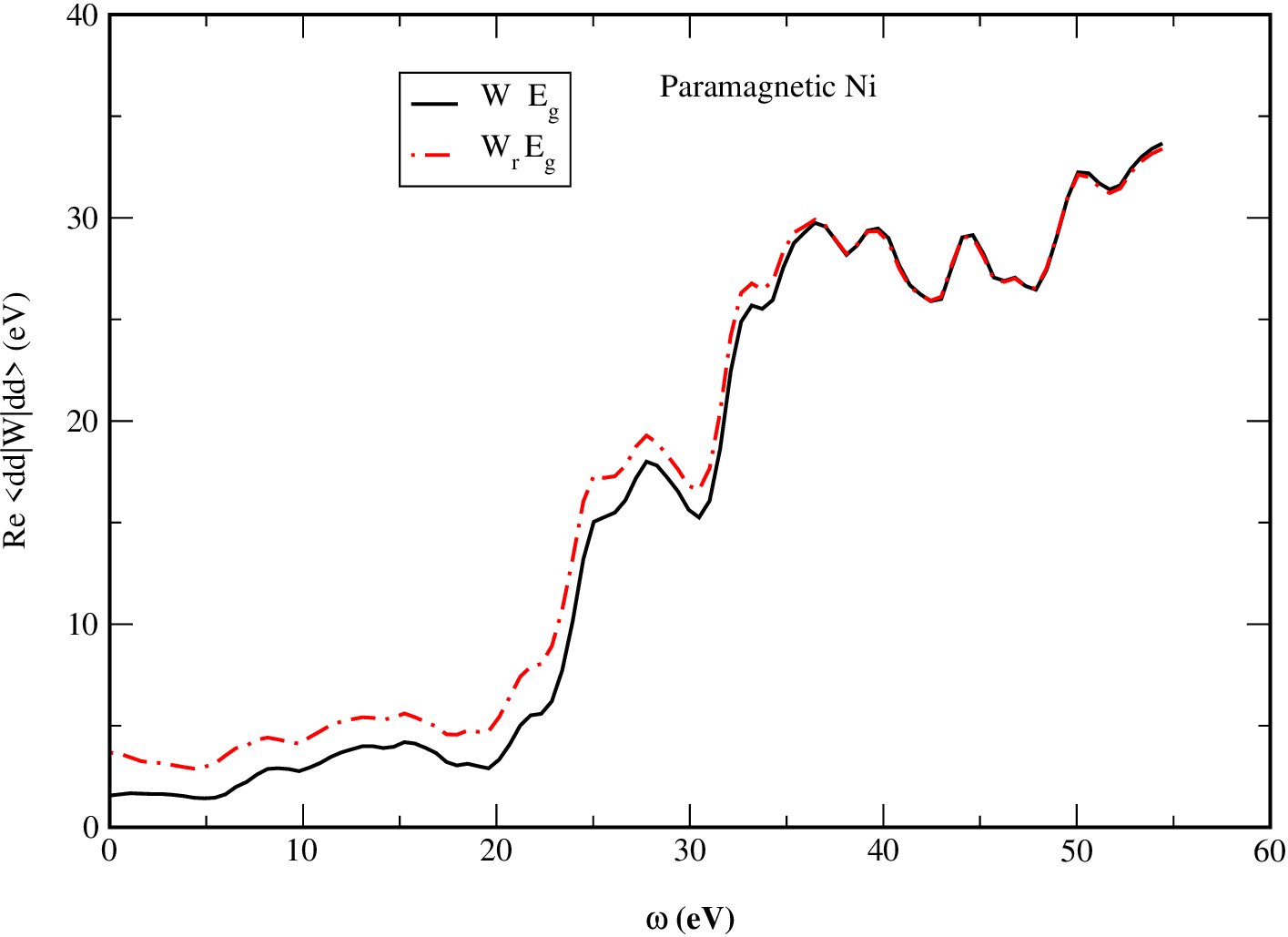}%
\caption{respect to the $\backslash$\$E$\backslash$\_g$\backslash$\$ channel.
Other channels give almost the same result.}%
\label{ReWr}%
\end{center}
\end{figure}
%EndExpansion

\section{Results and Discussions}

%AG
%AG:I HAVE MADE SOME SIGNIFICANT CHANGES TO THIS SECTION, MAINLY:
%
% -introduced subsections for clarity
%
% -reordered somewhat the topics
%
% -introduced additional explanations
%
%LETS HOPE IT IS ok W/ ALL OF US...
%
%

\subsection{Comparing self-energies}

The screened interaction with and without the 3d-3d transitions is shown in
Fig. (\ref{ReWr}) in the case of Nickel.
%AG
%
%
Here and in all the following, a spin-unpolarized (paramagnetic) solution is
considered. At low energies, the (partially) screened interaction $W_{r}$
without the 3d-3d transitions is larger than the full one $W$, and at high
energies they approach each other, as anticipated. Related calculations have
also been performed by Kotani~\cite{kotani}.

\begin{figure}
[ptb]
\begin{center}
\includegraphics[
height=2.2857in,
width=3.1566in
]%
{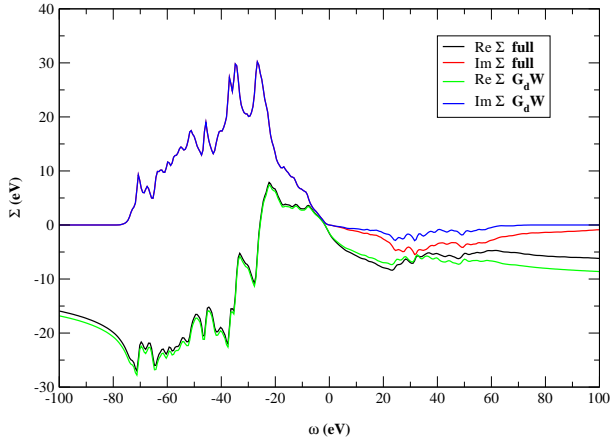}%
\caption{The self-energy of the real system (nickel) and the Hubbard model
with a frequency-dependent \emph{U.}}%
\label{Sig}%
\end{center}
\end{figure}

We first compare the self-energy obtained from a GW treatment of the full
system, given by:
\begin{equation}
\Sigma(r,r^{\prime};\omega)=\frac{i}{2\pi}\int d\omega^{\prime}e^{i\eta
\omega^{\prime}}G(r,r^{\prime};\omega+\omega^{\prime})W(r,r^{\prime}%
;\omega^{\prime}) \label{sigma_full}%
\end{equation}
to the self-energy obtained from the effective model (\ref{S}) with an
energy-dependent interaction $U(\omega)=W_{r}(\omega)$. Because of (\ref{W}),
the screened interaction corresponding to this $U(\omega)$ is simply $W$, and
the corresponding self-energy reads:
\begin{equation}
\Sigma_{d}(\omega)=\frac{i}{2\pi}\int d\omega^{\prime}e^{i\eta\omega^{\prime}%
}G_{d}(\omega+\omega^{\prime})W(\omega^{\prime}) \label{sigma_partial}%
\end{equation}
%AG added the following sentence:
%
%
The difference between this expression and the GWA for the full system
(Eq.\ref{sigma_full}) is that in (\ref{sigma_partial}) only the d-block of the
Green's function has been included (since the effective action was written for
the d-band only). Hence, the two self-energies differ by a term $G_{r} W$,
with $G_{r}=G-G_{d}$. We expect that the wavefunction overlap between two 3d
states (one from $G_{d}$ and the other from the 3d state appearing in the
matrix element of $\Sigma_{d}$) and other non-3d states is small so that
$\Sigma_{d}$ should be close to the true $\Sigma$. In Fig.~\ref{Sig}, the two
self-energies are displayed (more precisely, in this figure and in all the
following, we display the matrix element of the self-energy in the lowest 3d
state (band number 2), at the $\Gamma$-point, corresponding to an LDA
eigenenergy -1.79 eV). We observe that the two self-energies indeed almost
coincide with each other, even at high energies.
%AG
%
%
Hence, we conclude that the effective Hubbard model (\ref{S}) for the
d-subspace, with an energy-dependent interaction, provides a reliable
description of the real system.
%TCIMACRO{\FRAME{ftbpFU}{3.1566in}{2.2857in}{0pt}{\Qcb{The self-energy of the
%real system (nickel) and the Hubbard model with a frequency-dependent
%\emph{U.}}}{\Qlb{Sig}}{sigk0n2.eps}{\special{ language "Scientific Word";
%type "GRAPHIC";  maintain-aspect-ratio TRUE;  display "USEDEF";
%valid_file "F";  width 3.1566in;  height 2.2857in;  depth 0pt;
%original-width 9.4913in;  original-height 6.8571in;  cropleft "0";
%filename 'Sigk0n2.eps';file-properties "XNPEU";}}}%
%BeginExpansion
%EndExpansion

We now turn to the self-energy associated with the static Hubbard model
(\ref{ham}) with $U=W_{r}(0)\simeq3.5\mbox{eV}$.
%AG
% Ferdi: can you check that you agree w/ the value that I have put here, from your Fig.1:
%I think it helps to have the numerical value in the text.
%
%
%
The local screened interaction for this model, within the GW approximation, is
given by:
\begin{equation}
W_{d}(\omega)=[1-UP_{d}(\omega)]^{-1}U \label{Wd}%
\end{equation}
and the self-energy of the Hubbard model in the GWA thus reads:
\begin{equation}
\Sigma_{d}^{H}(r,r^{\prime};\omega)=\frac{i}{2\pi}\int d\omega^{\prime
}e^{i\eta\omega^{\prime}}G_{d}(r,r^{\prime};\omega+\omega^{\prime}%
)W_{d}(r,r^{\prime};\omega^{\prime}) \label{sigma_ham}%
\end{equation}
%
%As we mentioned above, if the Hubbard model and the assumption of a static
%\emph{U} is valid, the two self-energies or spectral functions should be close
%to each other, at least at low energies.%
%
%
%
%TCIMACRO{\FRAME{ftbpFU}{3.1531in}{2.2961in}{0pt}{\Qcb{The real parts of the
%self-energy of the real system (solid) and the Hubbard model (dash) in the
%GWA.}}{\Qlb{ReS}}{resigma2.eps}{\special{ language "Scientific Word";
%type "GRAPHIC";  maintain-aspect-ratio TRUE;  display "USEDEF";
%valid_file "F";  width 3.1531in;  height 2.2961in;  depth 0pt;
%original-width 9.436in;  original-height 6.8571in;  cropleft "0";
%filename 'ReSigma2.eps';file-properties "XNPEU";}}}%
%BeginExpansion
\begin{figure}
[ptb]
\begin{center}
\includegraphics[
height=2.2961in,
width=3.1531in
]%
{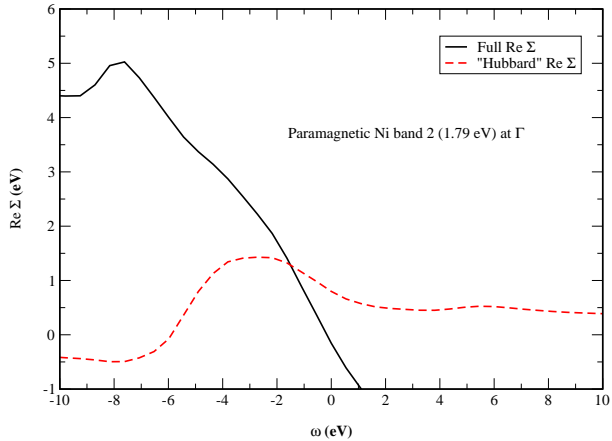}%
\caption{The real parts of the self-energy of the real system (solid) and the
Hubbard model (dash) in the GWA.}%
\label{ReS}%
\end{center}
\end{figure}
%EndExpansion
%
%
%
%AG
%
%
Note that the difference between this static Hubbard model self-energy and
that of the effective model with a frequency-dependent interaction (\ref{S})
relies in the use of a different form of the screened interaction $W_{d}$
instead of the full $W$. In Fig.~(\ref{ReS}), the real part of this
self-energy with that of the full GWA self-energy for nickel are shown. Since
the energy scale of the self-energy of the real system is determined by the
bare Coulomb interaction $v$ whereas the Hubbard self-energy is set by $U$,
the latter has been shifted so that it is equal to the former at the LDA
eigenvalue (-1.79 eV) of the band we have considered,
%AG I made a few changes below to be more precise: do you agree ?
%
%
%
at the $\Gamma$-point. The difference in magnitude of the self-energies is not
important since it simply shifts the spectrum
%AG
%
%
(or, said differently, we have compared differences $\Sigma-\mu$ from the
values of the chemical potential obtained in the various schemes). However,
the difference in the variation of the self-energy with respect to energy
matters, since it will give a different quasiparticle weight $Z=[1-\partial
\operatorname{Re}\Sigma/\partial\omega]^{-1}$ and affect the spectral
function. As can be deduced from the figure, the Z factor of the Hubbard model
taken at the energy of the quasiparticle band at the $\Gamma$-point
($\simeq-1.79$eV) is much closer to unity as compared to the true (full GW)
one because the former already contains the renormalization from the plasmon.
%AG
%
%
%
Hence, neglecting frequency-dependence directly affects the physical results,
\textit{even in the low-energy range}. We shall see below, however, that it is
possible to modify the static model in such a way that an accurate
approximation is obtained at low energy.
%TCIMACRO{\FRAME{ftbpFU}{3.192in}{2.3246in}{0pt}{\Qcb{The imaginary parts of
%the self-energy of the real systems (solid) and the Hubbard model (dash) in
%the GWA.}}{\Qlb{ImS}}{imsigma2.eps}{\special{ language "Scientific Word";
%type "GRAPHIC";  maintain-aspect-ratio TRUE;  display "USEDEF";
%valid_file "F";  width 3.192in;  height 2.3246in;  depth 0pt;
%original-width 9.436in;  original-height 6.8571in;  cropleft "0";
%filename 'ImSigma2.eps';file-properties "XNPEU";}}}%
%BeginExpansion
\begin{figure}
[ptbptb]
\begin{center}
\includegraphics[
height=2.3246in,
width=3.192in
]%
{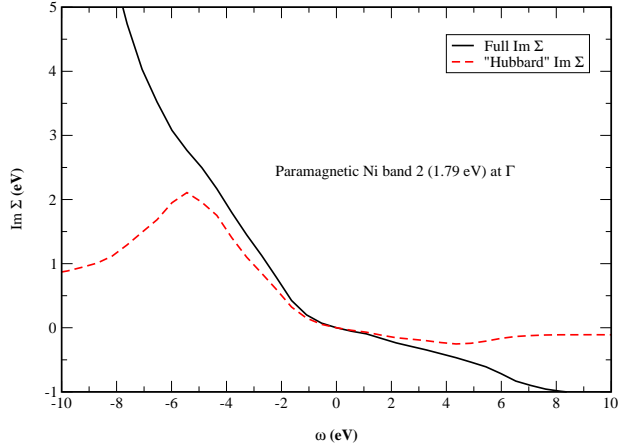}%
\caption{The imaginary parts of the self-energy of the real systems (solid)
and the Hubbard model (dash) in the GWA.}%
\label{ImS}%
\end{center}
\end{figure}
%EndExpansion

In Fig.~{\ref{ImS} the imaginary part of the self-energy is shown. Here we see
that Im $\Sigma_{d}^{H}$ for the Hubbard model is peaked around 5 eV since
there are no states above or below the 3d band. As a consequence, the real
part of the self-energy exhibits the Kramers-Kronig behavior at around -5 eV.
%AG
%
%
Within an energy region spanning about twice the 3d bandwidth, the imaginary
part of the Hubbard model self-energy, in contrast to the real part, is not in
bad agreement with the full one. }

%AG
% I added the following explanatory paragraph: do you agree ?
%
%
%
These findings can be understood qualitatively by considering more explicit
expressions of the self-energy obtained within the GW approximation, for an
effective model of the d-subspace defined on the periodic lattice. The
imaginary part reads:
\begin{equation}
\mbox{Im}\Sigma_{n}(k,\omega) = \sum_{q,m} \mbox{Im}W_{nm}(q,\omega
-\epsilon^{m}_{k-q})\, [n_{F}(\epsilon^{m}_{k-q})+n_{B}(\omega-\epsilon
^{m}_{k-q}] \label{formula_im}%
\end{equation}
In this expression, $n,m$ are band indices, $\epsilon^{n}_{k}$ corresponds to
the n-th non-interacting band and $n_{F}$ (resp. $n_{B}$) is the Fermi (resp.
Bose) function. $\mbox{Im}W$ is the spectral function associated with the
effective interaction (to be taken as $W$ if the self-energy of the
frequency-dependent effective model is considered, and as $W_{d}$ if that of
the static Hubbard model is considered). From this expression, one sees that
if one considers an energy $\omega$, the bands contained in the energy
interval $[\,0,|\omega|\,]$ are the only ones contributing significantly to
$\mbox{Im}\Sigma$. This is the reason why the \textit{imaginary part} of the
self-energy at low-energy will be correctly reproduced by the effective
low-energy model (provided of course that the spectral function $\mbox{Im}W$
is correctly approximated at low energy). In contrast, the real part of the
self-energy is obtained from the Kramers-Kr{\"o}nig relation in the form:
\begin{equation}
\mbox{Re}\Sigma_{n}(k,\omega) = -\frac{1}{\pi}\, P\int_{-\infty}^{+\infty}
d\nu\, \sum_{q,m} \mbox{Im}W_{nm}(q,\nu)\, \frac{n_{F}(\epsilon^{m}%
_{k-q})+n_{B}(\nu)}{\omega-\epsilon^{m}_{k-q}-\nu} \label{formula_re}%
\end{equation}
Because the principal- part integral extends over the whole frequency range,
high-frequency contributions influence the self-energy even at low frequency.
As a result, an accurate description of the real part of the self-energy
cannot be obtained within the naive construction of the static Hubbard model
because the effective interaction is not correctly approximated over the whole
frequency range. This formula also suggests a way to appropriately modify the
effective static model in order to obtain an accurate description at low
energy, as discussed below.

\subsection{The puzzle of the satellite}%

%TCIMACRO{\FRAME{ftbpFU}{3.1488in}{2.3065in}{0pt}{\Qcb{The spectral functions
%of the real system (solid) and the Hubbard model (dash) in the GWA.}%
%}{\Qlb{Akw}}{ak0n2.eps}{\special{ language "Scientific Word";
%type "GRAPHIC";  maintain-aspect-ratio TRUE;  display "USEDEF";
%valid_file "F";  width 3.1488in;  height 2.3065in;  depth 0pt;
%original-width 9.3789in;  original-height 6.8571in;  cropleft "0";
%filename 'Ak0n2.eps';file-properties "XNPEU";}}}%
%BeginExpansion
\begin{figure}
[ptb]
\begin{center}
\includegraphics[
height=2.3065in,
width=3.1488in
]%
{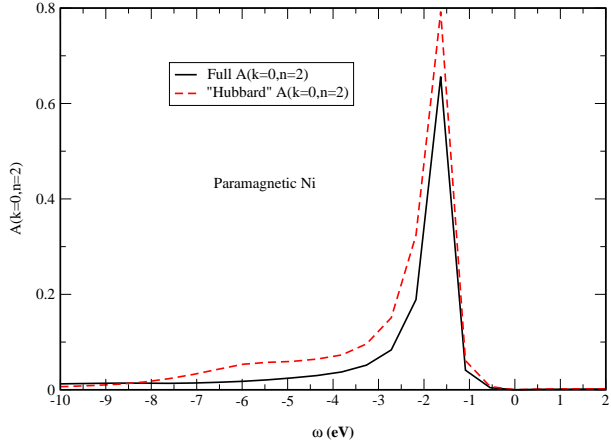}%
\caption{The spectral functions of the real system (solid) and the Hubbard
model (dash) in the GWA.}%
\label{Akw}%
\end{center}
\end{figure}
%EndExpansion
%
%
%
%AG
% I rewrote significantly the discussion of the satellite. In my view, whether
% the satellite obtained in the Hubbard model is spurious or not is quite
% an open question. Indeed, we know that the satellite is a true feature of the Hubbard
% model (eg when doing DMFT). On the other hand, it may be that using Hubbard is illegitimate
% in this energy range. I would suggest to leave this as an issue for further work...
%
%
%
Fig.~\ref{Akw} compares the spectral function obtained for the full system
within the GWA, and that of the static Hubbard model (\ref{ham}). A striking
difference between these two results is the absence of the 6 eV satellite in
the GWA for the full system, while the static Hubbard model displays a
satellite feature. This is to be expected from the structure of the
self-energy. The GWA self-energy of the full system continues to grow and
reaches a maximum around the plasmon excitation at around 25-30 eV, while the
self-energy of the static Hubbard model has a maximum around the width of the
3d band (4 eV). In comparison with the true plasmon, the \textquotedblright
plasmon\textquotedblright\ of the Hubbard model has a much lower energy, of
the order of $U$,
%AG
% In the original draft, U/v was written. I corrected it to U OK?
%
%
%
which results in the \textquotedblright6 eV satellite\textquotedblright. One
could blame the appearance of a satellite in our static Hubbard model
calculation on the fact that we have used a (non self-consistent) GW
approximation. However, more accurate treatments of the static Hubbard model
(such as DMFT) do preserve this feature, which has in this context a natural
interpretation as a lower Hubbard band. 
%On the other hand,
%sb 
%it could be that the experimentally observed Nickel satellite
%has a somewhat different physical origin and that the satellite obtained in
% FERDI, do you agree~?
%
%the physical origin of the experimentally observed Nickel satellite could be
%somewhat more complicated, so that the satellite obtained in 
%static Hubbard
%model calculations might be spurious in the context of Nickel (e.g because it
%is not legitimate to use a low-energy effective model in this energy range).
%In our view, this issue is an open problem which deserves further work.
%
On the other hand, it could be that
the experimentally observed Nickel satellite has a somewhat different physical
origin and that the satellite obtained in static Hubbard model calculations is
spurious in the context of Nickel (e.g because it is not legitimate to use a
low-energy effective model in this energy range). In our view, this issue is
an open problem which deserves further work.
%
%It should be noted, however, that the
%calculations have been performed for one iteration. It is conceivable that the
%satellite disappears or becomes smeared out at self-consistency, as found in
%the electron gas for the plasmon satellite. We should also keep in mind that
%we have employed the GWA in our investigation. It is not inconceivable that
%the GWA is not well suited for the Hubbard model in this range of couplings,
%but nevertheless we believe
%that the GWA illustrates the essential problem.
%
%

\subsection{Improving the effective static model}

The preceding discussion shows that an effective model for the d-band can be
constructed, which accurately reproduces the full results over an extended
energy range, provided the energy- dependence of $U(\omega)$ is retained.
However, performing calculations with an energy- dependent Hubbard interaction
is exceedingly difficult. A more modest goal is to obtain an effective model
which would apply to some low-energy range only (say, $|\omega|<\Lambda$, with
$\Lambda$ a cutoff of the order of the d-bandwith). In order to achieve this
goal, we propose to adopt a renormalization group point of view, in which high
energies are integrated out in a systematic way. Following this procedure, an
appropriate low-energy model with a static \emph{U} can be appropriately
constructed. As we shall see, the bare Green's function defining this
low-energy model does not coincide with the non-interacting Green's function
in the d-subspace (we have seen that this does not lead to a satisfactory
description, even at low energy).

Let us illustrate this idea within the GW approximation. The full Hilbert
space is divided into the Hubbard space, comprising the 3d orbitals, and the
downfolded space, comprising the rest of the Hilbert space. This approach is
complementary to the one in \cite{sun,silke} , where the division is done in
real space (onsite and off sites). Within the GWA we may write the full
self-energy as follows:
\begin{equation}
GW=G_{d}W_{d}+G_{d}(W-W_{d})+G_{r}W. \label{GW}%
\end{equation}
where $G=G_{d}+G_{r}$. The term $G_{d}(W-W_{d})$ represents the high-energy
contribution of the screened interaction. This is the main source of error in
the naive static limit, as discussed in the previous sections. The term
$G_{r}W$ is \emph{not} obtainable within the Hubbard model, even when a
frequency-dependent \emph{U} is employed since $G_{r}$ resides in the
downfolded space. This term was shown to be small. Its effect at low-energy
can also be taken into account by appropriate modifications of the
one-particle propagator, but we shall neglect it for simplicity.

We consider the following Hubbard model with a static $U=W_{r}(0)$, but a
modified one-particle propagator $\widetilde{G}_{0}$, defined by the action:
\begin{align}
S_{H} &  =-\int d\tau d\tau^{\prime}\sum d_{Rn}^{\dagger}(\tau)[\widetilde
{G}_{Rn,R^{\prime}n^{\prime}}^{0}]^{-1}(\tau-\tau^{\prime})d_{Rn^{\prime}%
}(\tau^{\prime})\nonumber\\
&  +\frac{1}{2}\int d\tau\sum:d_{R_{1}n}^{\dagger}(\tau)d_{R_{2}n^{\prime}%
}(\tau):U_{R_{1}nR_{2}n^{\prime},R_{3}mR_{4}m^{\prime}}:d_{R_{3}m}^{\dagger
}(\tau)d_{R_{4}m^{\prime}}(\tau):\label{SH}%
\end{align}
The self-energy of this static Hubbard model in the GWA is:
\begin{equation}
\widetilde{\Sigma}_{d}^{H}=\widetilde{G}_{d}\widetilde{W}_{d}%
\end{equation}
where the new effective interaction is $\widetilde{W}_{d}=U\,[1-U\widetilde
{P}_{d}]^{-1}$, with $\widetilde{P}_{d}$ constructed from the new Green's
functions $\widetilde{G}_{d}$ (schematically $\widetilde{P}_{d}=\widetilde
{G}_{d}\cdot\widetilde{G}_{d}$). We request that the interacting Green's
function of this modified static model coincides with that of the Green's
function calculated with the frequency- dependent interaction \textit{in the
low energy range} $|\omega|<\Lambda$:), that is:
\begin{equation}
\widetilde{G}_{d}^{-1}-\widetilde{\Sigma}_{d}^{H}\simeq G_{d}^{-1}-G_{d}%
W_{d}-G_{d}(W-W_{d})\,\,\,\mbox{for}\,\,|\omega|<\Lambda
\end{equation}
Using the identity $G^{-1}-GW=[1-UP]^{-1}\,G^{-1}$, this can be rewritten as:
\begin{equation}
\lbrack1-U\widetilde{P}_{d}]^{-1}\,\widetilde{G}_{d}^{-1}\simeq\lbrack
1-UP_{d}]^{-1}\,G_{d}^{-1}-G_{d}(W-W_{d})\,\,\,\mbox{for}\,\,|\omega|<\Lambda
\end{equation}
This is an integral equation which determines in principle the modified bare
Green's function $\widetilde{G}_{d}$ to be used in the ''downfolded'' static
action (\ref{SH}). To first approximation, one can neglect the polarization
terms in this equation, and obtain the first order modification of
$\widetilde{G}_{d}$ as:
\begin{equation}
\widetilde{G}_{d}^{-1}=G_{d}^{-1}-G_{d}(W-W_{d})+\cdots
\end{equation}
The first correction appearing in this equation is precisely the contribution
coming from the high-energy part of the screened interaction. We have
explained above that this correction is not small, which is the reason for the
failure of the ''naive'' static Hubbard model using the non-interacting
$G_{d}$. Dividing the screened interaction into a low-energy part for
$|\omega|<\Lambda$ and a high-energy part for $|\omega|>\Lambda$, we can use
the explicit forms (\ref{formula_im},\ref{formula_re}) given in the previous
section to obtain this first correction in the form:
\begin{equation}
\widetilde{G}_{d}^{-1}(k,\omega)_{n}=G_{d}^{-1}(k,\omega)_{n}+\frac{1}{\pi
}\int_{|\nu|>\Lambda}d\nu\,\sum_{q,m}\mbox{Im}W_{nm}(q,\nu)\,\frac
{n_{F}(\epsilon_{k-q}^{m})+n_{B}(\nu)}{\omega-\epsilon_{k-q}^{m}-\nu}%
+\cdots\label{correction}%
\end{equation}
In particular, we see by expanding this expression to first order in $\omega$,
that the low frequency expansion of the modified one-particle propagator
reads: $\widetilde{G}_{d}^{-1}(k,\omega+i0^{+})_{n}=(1+\alpha_{k,n}%
)\omega-\epsilon_{k}^{n}+\cdots$. The coefficient $\alpha_{k,n}$ is a partial
contribution to the quasi-particle residue $\alpha_{k,n}=-\frac{1}{\pi}%
\sum_{m}\int_{|\nu|>\Lambda}d\nu\,\mbox{Im}W_{nm}(q,\nu)\,\frac{n_{F}%
(\epsilon_{k-q}^{m})+n_{B}(\nu)}{(\nu-\epsilon_{k-q}^{m})^{2}}$. In practice
this integral can be easily evaluated using the well known plasmon pole
approximation, which should contain most of the high energy contribution. This
correction insures that the quasi- particle residue is obtained correctly from
this ``downfolded'' static model. We do not, however, expect that this
improvement solves the problem with the satellite discussed in the previous subsection.

\subsection{Beyond the GW approximation}

In strongly correlated systems, it is known that the GWA is not sufficient and
improvement beyond the GWA is needed. Much of the short coming of the GWA
probably originates from the improper treatment of \textit{short-range}
correlations within the random-phase approximation (RPA). Thus, one would like
in the first instance attempt to improve these short-range correlations, which
are essentially captured by the Hubbard model. The formula (\ref{GW}) suggests
a natural way to do this, as follows. The contribution from the
frequency-dependent \emph{U }as well as the contribution from the downfolded
Hilbert space are treated within the GWA. The self-energy $G_{d}W_{d}$
corresponding to the \emph{GW} self-energy of the Hubbard model with a static
\emph{U }can then be replaced by that obtained from more accurate theories
such as dynamical mean-field theory (DMFT) \cite{georges} or from exact
methods such as Lanczos diagonalization. Thus, if we use the LDA to construct
$G_{d}$, the correction to the LDA exchange-correlation potential reads:
\begin{equation}
\Delta\Sigma=\Sigma_{H}+G_{d}(W-W_{d})+G_{r}W-v_{xc}, \label{dSigma}%
\end{equation}
where $\Sigma_{H}$ is the Hubbard model self-energy obtained from more
accurate methods, replacing $G_{d}W_{d}$, We note also that the scheme in
(\ref{GW}) avoids the problem of double counting, inherent in LDA+U
\cite{anisimov} or LDA+DMFT methods.\cite{anisimov2} In this way, it is
possible to calculate the Hubbard \emph{U} using the response function
constructed from the LDA bandstructure.

The application of DMFT requires the Hubbard \emph{U} for the impurity model.
To calculate the Hubbard \emph{U} for an impurity model, it is necessary to
downfold contributions to the polarization \emph{P} from the neighboring
sites. Here, the identity in Eq. (\ref{W}) shows its usefullness. Since the
formulation in (\ref{W}) is quite general, we merely need to redefine $P_{d}$
to be the onsite polarizations or equivalently, we redefine $P_{r}$ to also
include polarizations from the neighboring sites. The computational result for
Ni shows that the difference between the lattice and the impurity Hubbard
\emph{U} is rather small, essentially negligible. Somewhat different but
related calculations of the Hubbard \emph{U} for the impurity model has also
been performed in \cite{kotani} for Fe and Ni, which confirm the result
obtained in our formulation.

Another feasible approach for calculating physical quantities of the derived
Hubbard model is by using the path integral renormalization group (PIRG)
method.\cite{imada} One advantage of this method is the possibility of using
the lattice Hubbard model as opposed to the impurity model, thus including
spatial fluctuations and possible symmetry breaking. The method is also suited
for studying single-particle Green's functions as well as thermodynamic
quantities, in particular for accurate determination of the phase diagram,
which often requires a determination of the possible symmetry breaking of the
Hubbard Hamiltonian after taking account of spatial and temporal fluctuations
on an equal footing. In calculating the self-energy, one may substitute
$G_{d}W_{d}$ by the self-energy obtained within the correlator projection
method \cite{onoda} together with the PIRG. These form a future challenge in
the field of strongly correlated materials.

%AG
%
%

\section{Summary and conclusion}

In conclusion, we have investigated the construction of effective models for
the correlated orbitals in materials with a natural separation of bands. We
have shown that, if one retains the full frequency dependence of the local
components of the screened interaction, an accurate effective model can be
obtained over an extended energy range. Simply neglecting this energy
dependence and using the non-interacting hamiltonian into a static Hubbard
model does not provide an accurate description even at low energy. However, a
proper modification of the bare propagator, obtained by integrating out high
energies, allows for the construction of an effective Hubbard model which
describes the low-energy physics in a satisfactory manner.

\section{Acknowledgment}

We thank Y. Asai for useful comments.
%AG
%
%
We would like to acknowledge hospitality of the Kavli Institute for
Theoretical Physics (KITP at UC-Santa Barbara) where this work was initiated
and partially supported by grant NSF-PHY99-07949. G.K aknowledges NSF grant
DMR-0096462 and F.A acknowledges the support from NAREGI Nanoscience Project,
Ministry of Education, Culture, Sports, Science and Technology, Japan. We also
acknowledge the support of an international collaborative grant from CNRS
(PICS 1062), of an RTN network of the European Union and of IDRIS Orsay
(project number 031393).

\end{document}